# Acoustic waveguide filters made up of rigid stacked materials with elastic joints

Andrea Bacigalupo[1*], Luigi Gambarotta[2], Marco Lepidi[2], Francesca Vadalà[2]
[1]IMT School for Advanced Studies, Lucca, Italy
[2]Department of Civil, Chemical and Environmental Engineering, University of Genova, Italy

**Abstract**

The acoustic dispersion properties of monodimensional waveguide filters can be assessed by means of the simple prototypical mechanical system made of an infinite stack of periodic massive blocks, connected to each other by elastic joints. The linear undamped dynamics of the periodic cell is governed by a two degree-of-freedom Lagrangian model. The eigenproblem governing the free propagation of shear and moment waves is solved analytically and the two dispersion relations are obtained in a suited closed form fashion. Therefore, the pass and stop bandwidths are conveniently determined in the minimal space of the independent mechanical parameters. Stop bands in the ultra-low frequency range are achieved by coupling the stacked material with an elastic half-space modelled as a Winkler support. A convenient fine approximation of the dispersion relations is pursued by formulating homogenised micropolar continuum models. An enhanced continualization approach, employing a proper Maclaurin approximation of pseudo-differential operators, is adopted to successfully approximate the acoustic and optical branches of the dispersion spectrum of the Lagrangian models, both in the absence and in the presence of the elastic support.

## 1. Introduction

The design of acoustic filters to control the attenuation and isolation of vibrations is mostly based on engineered materials and devices with periodic structure (see for reference Vasseur *et al.*, 2001, Deymier, 2013, Ma and Sheng, 2016, Habermann and Norris, 2016, Cummer *et al.*, 2016). Of particular focus is the design of filters exhibiting complete band gaps, with wave propagation prevented regardless of mode and wave vector, or the design of acoustic waveguides. As it is well known, band gaps in periodic media may take place as effects of Bragg scattering phenomena, resulting from the destructive interference of the wave reflection from the periodic components within the medium (Deymier, 2013, Hussein *et al.*, 2014). However, it must be emphasized that in most technical applications low frequency band gaps are required, which may hardly be obtained from the Bragg scattering. In fact, to obtain low-frequency Bragg gaps, heavy inclusions, low stiffness and large cell size are needed, which may be not suitable for practical purposes. Alternatively, band gaps may be obtained by including local resonators in the periodic material, whose role is to impede wave propagation around their resonance frequency by transferring the vibrational energy to the resonator (see for reference Huang and Sun, 2010, Craster and Guenneau, 2012, Baravelli and Ruzzene, 2013, Bacigalupo and Gambarotta, 2017, Beli and Ruzzene, 2018, Deng *et al.*, 2018). In this case, low-frequency band gaps may be obtained, even if heavy resonators and low stiffness cells are required to get wide stop-bands. Improvements in creating wide low frequency band gaps have been obtained by virtue of the concept of effective inertia, consisting in amplifying a small mass through embedded amplification mechanisms (see Yilmaz *et al.*, 2007, Acar and Yilmaz, 2013, Taniker and Yilmaz, 2015, Frandsen *et al.*, 2016) or by mimicking biological structural systems (see Miniaci *et al.*, 2016 and Miniaci *et al.*, 2018). Recently, Yin et al., 2014, 2015, and Chen and Wang, 2015, obtained ultra-wide low-frequency band gaps in composites designed on the thought of staggered and combined soft and hard materials. These phononic crystals mimic the microstructure of the nacre and may be idealized as an assemblage of rigid blocks and elastic interfaces (see also Matlack et al., 2018).

---

[*] Corresponding Author

In this paper the acoustic properties of a stack of rigid rectangular blocks connected through elastic interfaces is analyzed, in order to understand the geometric and constitutive conditions for which interesting dispersive properties are attained. Accordingly, the analysis outcomes apply to all the technological realizations for which the mechanical assumption of linear elastic interfaces is justified. The wave modes are described in terms of transverse displacement and rotation of the blocks and the acoustic spectrum is characterized by two dispersion functions. The analytical model allows the analysis of the influence of the block geometry and of the constitutive parameters of the interfaces on the conditions under which low frequency band gaps may be attained. To improve the acoustic performances of the blocky system and to achieve ultra-low frequency stop bands, the blocks have been coupled with an elastic half-space model as a Winkler support. In this case, the low frequency branch has a non-vanishing frequency at the long wave-length limit and the opening of a ultra-low frequency band gap may be obtained (see Gei *et al.*, 2009, Brun *et al.*, 2012, Piccolroaz and Movchan, 2014, and Carta *et al.*, 2017). Moreover, a convenient fine approximation of the dispersion functions of the Lagrangian model is pursued through the formulation of an equivalent continuum model. This continuum model is based on an enhanced continualization approach for discrete models proposed by Bacigalupo and Gambarotta, 2018 in order to circumvent some drawbacks emerging by applying classical continualization approaches (Bacigalupo and Gambarotta, 2017). This procedure provides different equivalent continua with increasing order of the leading derivative with non-local constitutive and inertial terms. Finally, the capabilities and the validity limits of the enhanced continuum are emphasized through the approximation of the dispersive functions of the discrete model for selected values of the geometric and constitutive parameters.

## 2. Wave propagation in a rigid stacked material with elastic interfaces

Let us consider a linear assemblage of rigid blocks of length $\ell$ and width $b$ connected to each other by elastic interfaces having normal stiffness $k_n$ and tangential stiffness $k_t$, respectively (see Figure 1). The blocks have unit thickness and mass density $\rho$.

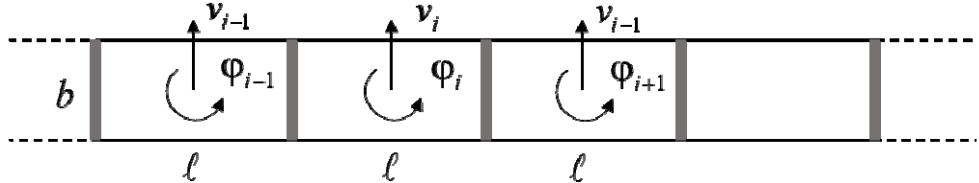

Fig. 1. The stack of rigid blocks and the generalized displacements

According to a Lagrangian mechanical formulation, the *i*-th block is assumed to undergo a transverse displacement $v_i$ and rotation $\varphi_i$ so that the equations of free motion are written (see for reference Bacigalupo and Gambarotta, 2017)

$$\begin{cases} k_t b \left( v_{i+1} - 2v_i + v_{i-1} \right) - \frac{1}{2} k_t b \ell \left( \varphi_{i+1} - \varphi_{i-1} \right) - M \ddot{v}_i = 0 \\ \frac{1}{2} k_t b \ell \left( v_{i+1} - v_{i-1} \right) + \frac{1}{12} k_n b^3 \left( \varphi_{i+1} - 2\varphi_i + \varphi_{i-1} \right) + \\ -\frac{1}{4} k_t b \ell^2 \left( \varphi_{i+1} + 2\varphi_i + \varphi_{i-1} \right) - J \ddot{\varphi}_i = 0 \end{cases}, \qquad (1)$$

where $k_t b$ and $\frac{1}{12} k_n b^3$ are the shear and flexural stiffness of the interface, respectively, and $M = \rho b \ell$ and $J = \frac{1}{12} M \left( b^2 + \ell^2 \right)$ the mass and the rotational inertia of the block. The displacement field is represented through a non-dimensional variable $\psi_i = \frac{v_i}{\ell}$ and the equation of motion may be written in the space non-dimensional form

$$\begin{cases} \left( \psi_{i+1} - 2\psi_i + \psi_{i-1} \right) - \frac{1}{2} \left( \varphi_{i+1} - \varphi_{i-1} \right) = I_\psi \ddot{\psi}_i \\ \frac{1}{2} \left( \psi_{i+1} - \psi_{i-1} \right) + \frac{1}{12} r_k r_b^2 \left( \varphi_{i+1} - 2\varphi_i + \varphi_{i-1} \right) + \\ -\frac{1}{4} \left( \varphi_{i+1} + 2\varphi_i + \varphi_{i-1} \right) = \frac{1}{12} I_\psi \left( 1 + r_b^2 \right) \ddot{\varphi}_i \end{cases}, \qquad (2)$$

having introduced the term $I_\psi = \frac{M}{k_t b} = \frac{\rho \ell}{k_t}$, the constitutive ratio $r_k = \frac{k_n}{k_t}$ and the geometric ratio $r_b = \frac{b}{\ell}$, respectively.

Elastic transverse and rotational waves are analysed in the canonical form with the generalized displacement of the $i$-th node written in the form $\psi_i(t) = \bar{\psi} \exp\left[ I \left( k x_i - \omega t \right) \right]$, $\varphi_i(t) = \bar{\varphi} \exp\left[ I \left( k x_i - \omega t \right) \right]$, being $I$ the imaginary unit, $k$ the wave number, $\omega$ the circular frequency and $\bar{\psi}$ and $\bar{\varphi}$ the wave amplitudes. Accordingly, the equations governing the propagation of harmonic waves are

$$\mathbf{H}(k\ell, \omega) \mathbf{\upsilon} = \left( \mathbf{H}_{Lag} - \omega^2 \mathbf{I}_{Lag} \right) \mathbf{\upsilon} = \mathbf{0}, \qquad (3)$$

where $\mathbf{H}_{Lag}$ is a Hermitian matrix and $\mathbf{I}_{Lag}$ is a diagonal matrix assuming the form

$$\mathbf{H}_{Lag} = \begin{bmatrix} 2\left[ 1 - \cos(k\ell) \right] & I \sin(k\ell) \\ -I \sin(k\ell) & \frac{1}{2}\left[ \left( 1 + \frac{1}{3} r_k r_b^2 \right) + \left( 1 - \frac{1}{3} r_k r_b^2 \right) \cos(k\ell) \right] \end{bmatrix},$$

$$\mathbf{I}_{Lag} = I_\psi \begin{bmatrix} 1 & 0 \\ 0 & \frac{1}{12}\left( 1 + r_b^2 \right) \end{bmatrix} \qquad (4)$$

and $\mathbf{\upsilon}^T = \{ \bar{\psi} \quad \bar{\varphi} \}$ is the polarization vector. As usual, by solving the characteristic equation $\det \mathbf{H}(k\ell, \omega) = 0$, two dispersion functions $\omega(k\ell)$ are obtained, which depend on both the geometric ratio $r_b$ and the constitutive terms $r_k$, $I_\psi$.

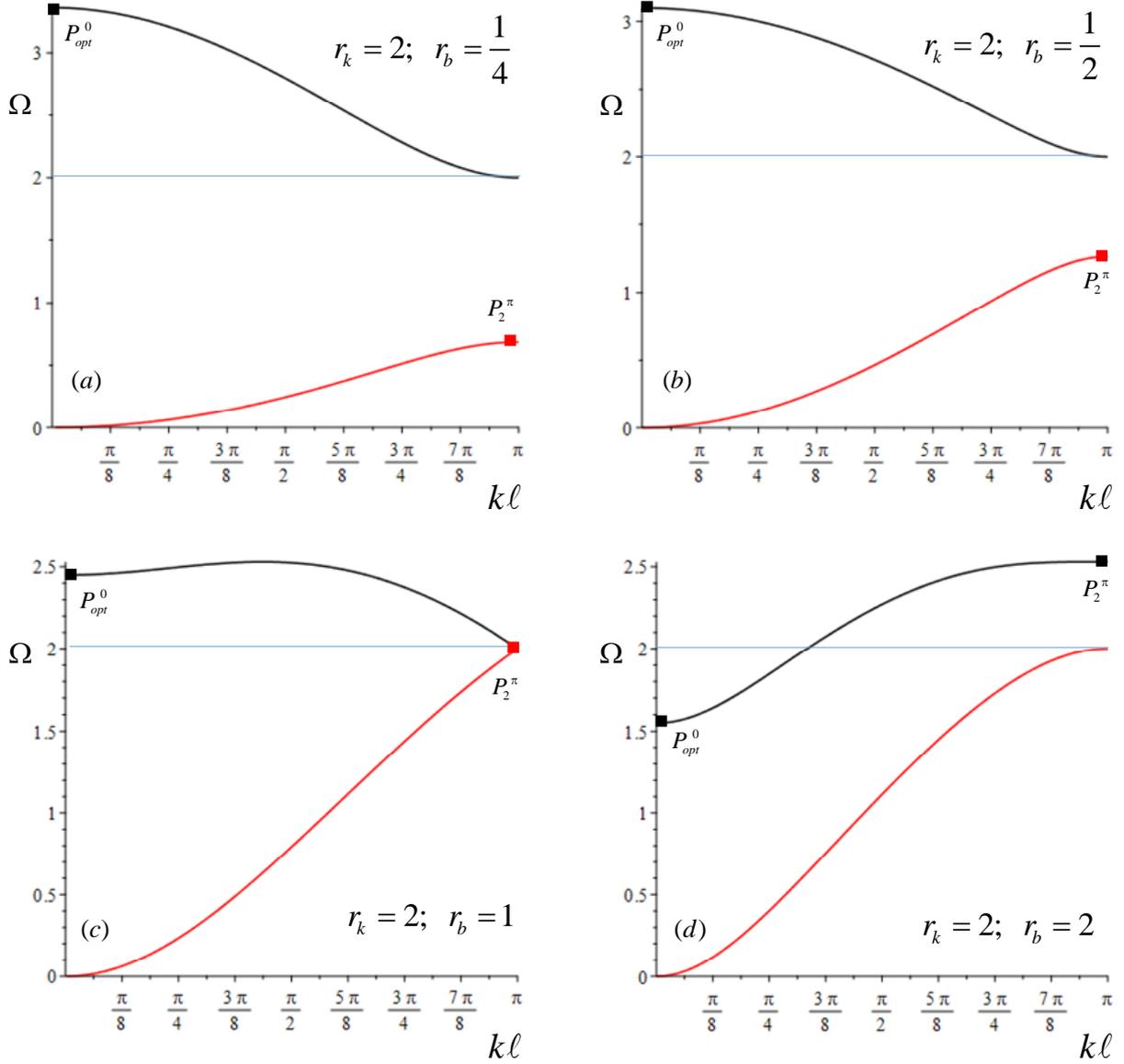

Fig. 2. Floquet-Bloch spectra for $r_k = 2$ with transition from existence of a stop band ($r_b < 1$) to a pass band ($r_b \geq 1$).

Tracking the dependence on the wavenumber $k$ in the domain $[0, \pi]$ allows to completely assess the Floquet-Bloch spectrum of the periodic system (Brillouin, 1946). The dispersion function may be represented in terms of the non-dimensional circular frequency $\Omega = \omega \sqrt{I_\psi}$, so that the acoustic and optical branches of the spectrum are defined by the relations $\Omega_{ac}(k\ell)$ and $\Omega_{opt}(k\ell)$. In the long wave-length regime $(k\ell \to 0)$ the non-dimensional frequencies are $\Omega_{ac}^0 = 0$ and $\Omega_{opt}^0 = \dfrac{2\sqrt{3}}{\sqrt{1+r_b^2}}$. The first frequency identifies the starting point of the acoustic branch and is systematically null. The second frequency identifies the starting point of the optical branch and does not depend on the constitutive ratio. In the short wave-length regime $(k\ell = \pi)$, the two non-dimensional frequencies attain the values $\Omega_1^\pi = 2$ and $\Omega_2^\pi = 2r_b\sqrt{\dfrac{r_k}{1+r_b^2}}$, which cannot be identified *a priori* as limit points of the acoustic or optical branch. However, the occurrence of the frequency coalescence $\Omega_1^\pi = \Omega_2^\pi$, denoting the closing of a band gap for short waves and in general in the Bloch spectrum is obtained for the particular parameter combination $r_k = \dfrac{1+r_b^2}{r_b^2}$. If the constitutive ratio is fixed to the realistic value $r_k = 2$, the coalescence condition is satisfied by square blocks ($r_b = 1$).

Tab. 1. Limit points of the acoustic and optical branches for $r_k = 2$.

|  | $\Omega_{opt}^0 = \dfrac{2\sqrt{3}}{\sqrt{1+r_b^2}}$ | $\Omega_2^\pi = 2r_b\sqrt{\dfrac{r_k}{1+r_b^2}}$ |
|---|---|---|
| $r_b = \dfrac{1}{4}$ | $P_{opt}^0 = \left(0, \dfrac{8\sqrt{51}}{17}\right)$ | $P_2^\pi = \left(0, \dfrac{2\sqrt{34}}{17}\right)$ |
| $r_b = \dfrac{1}{2}$ | $P_{opt}^0 = \left(0, \dfrac{4\sqrt{15}}{5}\right)$ | $P_2^\pi = \left(0, \dfrac{2\sqrt{10}}{5}\right)$ |
| $r_b = 1$ | $P_{opt}^0 = \left(0, \sqrt{6}\right)$ | $P_2^\pi = (0, 2)$ |
| $r_b = 2$ | $P_{opt}^0 = \left(0, \dfrac{2\sqrt{15}}{5}\right)$ | $P_2^\pi = \left(0, \dfrac{4\sqrt{10}}{5}\right)$ |

The Floquet-Bloch spectra for different geometric ratios are shown in Figure 2 for a fixed value of the constitutive ratio. Some typical qualitative scenarios are portrayed, corresponding to the existence of a frequency stop band for rectangular blocks with $b < \ell$ ($r_b < 1$ in Figure 2a,b), the closing of the band gap due to the frequency coalescence for square blocks ($r_b = 1$ in Figure 2c) and, finally, the absence of stop bands for rectangular blocks with $b > \ell$ ($r_b > 1$ in Figure 2d). The frequency values attained by the acoustic and optical branches at the limit of short and long wavelengths (marked by the points $P_{opt}^0, P_2^\pi$ in Figure 2) are reported in Table 1. In particular, the frequency at the starting point of the optical branch $\Omega_{opt}^0$ is found to decrease monotonically with increasing geometric ratio $r_b$. On the contrary, the frequency $\Omega_2^\pi$ is found to increase monotonically with increasing geometric ratio $r_b$.

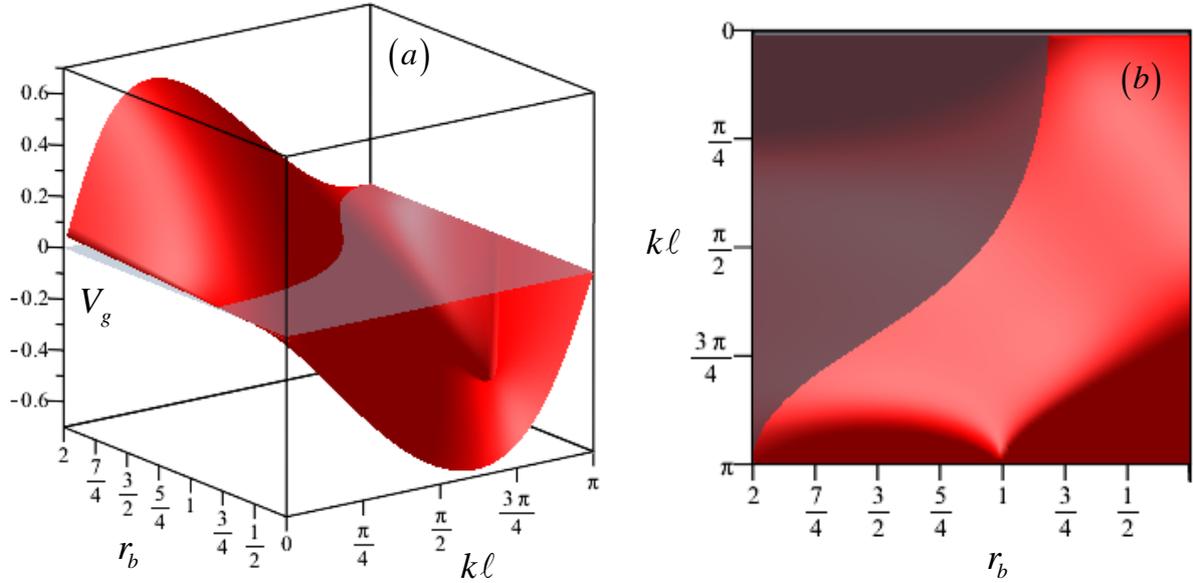

Fig.3. Non-dimensional group velocity $V_g$ associated to the optical branch for $r_k = 2$ and varying $r_b$ and $k\ell$. (a) 3D view; (b) projection on the $(r_b, k\ell)$ plane (dark grey – positive group velocity, red – negative velocity).

From Figure 2, the dispersion curves show that the group velocity associated to the optical branch may be negative. This circumstance is highlighted in the diagrams of Figure 3, where the non-dimensional group velocity $V_g = \frac{\sqrt{I_\psi}}{\ell} \frac{\partial \omega}{\partial k} = \frac{\partial \Omega}{\partial (k\ell)}$ is shown in terms of the geometric ratio $r_b \in [1/4, 2]$ and the non-dimensional wave number $k\ell \in [0, \pi]$ for the realistic constitutive ratio $r_k = 2$.

In order to evaluate the performance of the periodic system as acoustic filter, it is interesting to assess the stop band amplitude $A_s(r_b, r_k)$, which can be mathematically defined in the form

$$A_s(r_b, r_k) = \frac{1}{2}\left(\left|\min_{0 \leq kl \leq \pi} \Omega_{opt} - \max_{0 \leq kl \leq \pi} \Omega_{ac}\right| + \min_{0 \leq kl \leq \pi} \Omega_{opt} - \max_{0 \leq kl \leq \pi} \Omega_{ac}\right). \tag{5}$$

The minimum and the maximum values of the optical and acoustic branches can be expressed in terms of the mechanical parameters $r_b$ and $r_k$

$$\min_{0 \leq kl \leq \pi} \Omega_{opt} = \min\left\{\frac{2\sqrt{3}}{\sqrt{r_b^2+1}}; \sqrt{\frac{1+(r_k+1)r_b^2+C(r_k,r_b)}{r_b^2+1}}\right\},$$
$$\max_{0 \leq kl \leq \pi} \Omega_{ac} = \sqrt{\frac{2+2(r_k+1)r_b^2+2C(r_k,r_b)}{r_b^2+1}}. \tag{6}$$

where the auxiliary function $C(r_k, r_b) = ((r_k-1)r_b^2 - 1)\text{csgn}((r_k-1)r_b^2 - 1)$. The stop band amplitude $A_s(r_b, r_k)$ is shown in Figure 4 for the relevant ranges of the two parameters. It is worth noting that maximal amplitude is reached for low geometric ratios in combination with small values of the constitutive ratio (Figure 4a). The band gap is absent for the large values of the geometric ratios corresponding to the white region in Figure 4b.

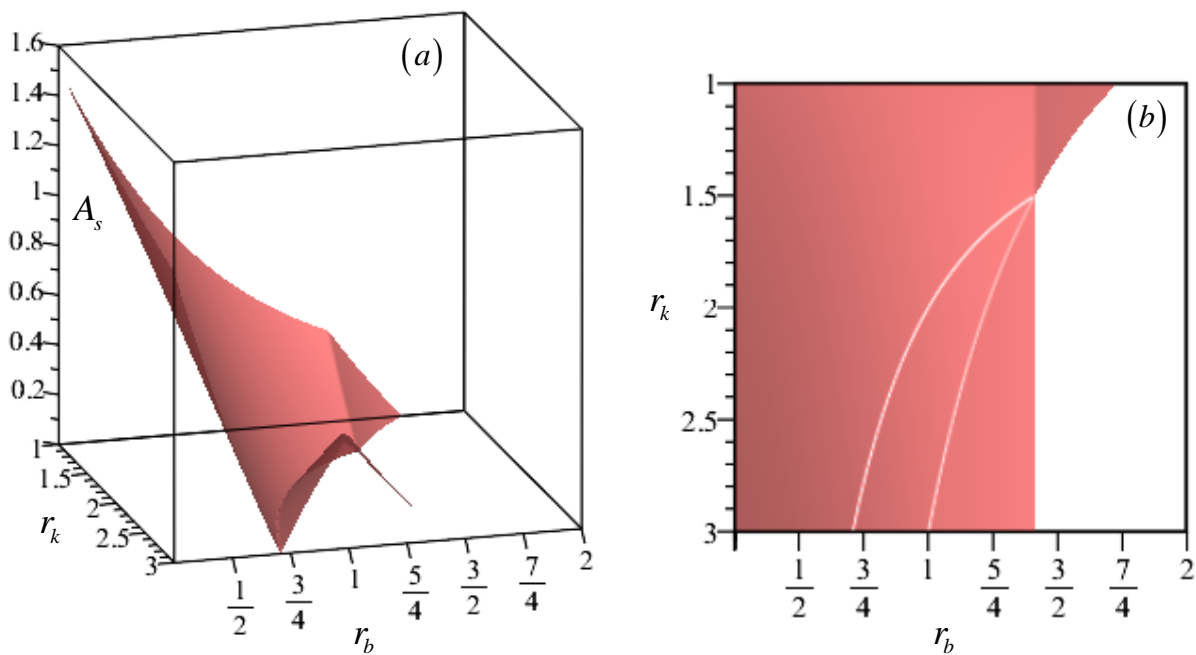

Fig. 4. Influence of the model parameters on the opening of the stop band amplitude.

In order to evaluate the performance of the periodic system as acoustic waveguide filter, it is also interesting to assess the total amplitude $A_p(r_b, r_k)$ of the pass bands in the spectrum, which can be mathematically defined in the form

$$A_p(r_b, r_k) = \left(\max_{0 \leq kl \leq \pi} \Omega_{opt} - \min_{0 \leq kl \leq \pi} \Omega_{ac}\right) - A_s(r_b, r_k), \qquad (7)$$

where the maximum and the minimum values of the optical and acoustic branch expressed in terms of the mechanical parameters $r_b$ and $r_k$ are

$$\max_{0 \leq kl \leq \pi} \Omega_{opt} = \max\left\{\frac{2\sqrt{3}}{\sqrt{r_b^2+1}}; \sqrt{\frac{1+(r_k+1)r_b^2 + C(r_k, r_b)}{r_b^2+1}}\right\},$$

$$\min_{0 \leq kl \leq \pi} \Omega_{ac} = 0. \qquad (8)$$

The total amplitude $A_p(r_b, r_k)$ of the pass bands is shown in Figure 5 (green-yellow manifold) for the relevant ranges of the two parameters. The maxima of the amplitude tend to occur for large constitutive ratios. The stop band amplitude $A_s(r_b, r_k)$ is also shown (red manifold) for comparison (Figure 5a). It is worth remarking that the stop band co-exists with a pass band, whereas the pass band exists in the absence of stop band (green-yellow region in Figure 5b). If the pass and the stop band co-exist, the pass bandwidth is systematically larger than the stop bandwidth. Comparable amplitudes occur for low geometric ratios in combination with small values of the constitutive ratio (Figure 5a).

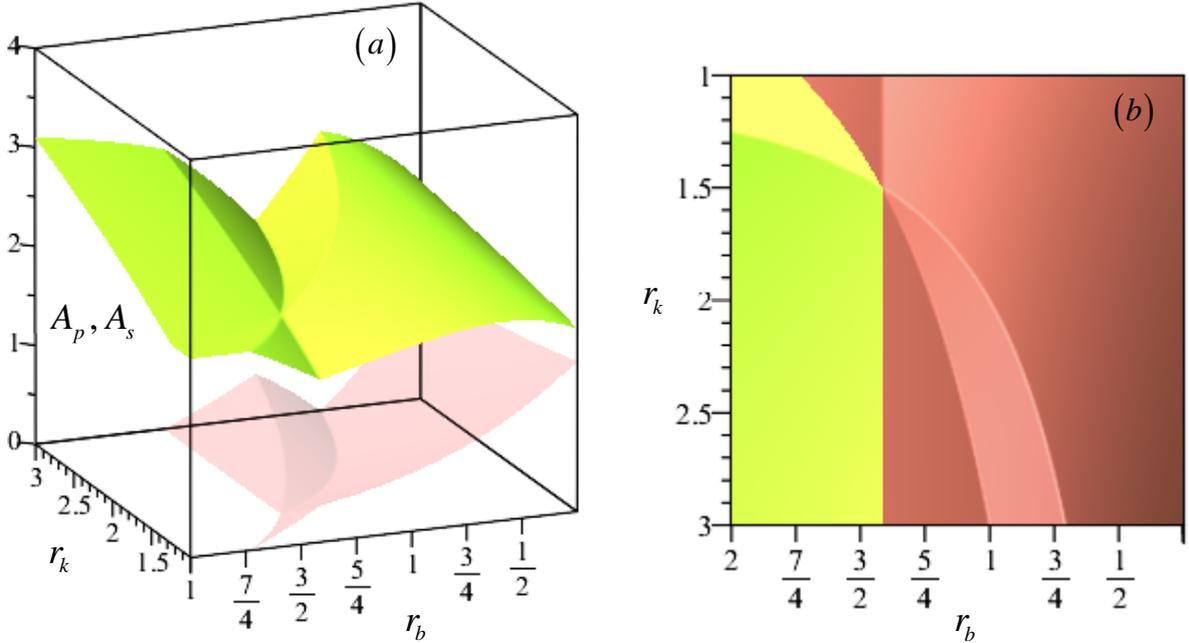

Fig. 5. Influence of the model parameters on the pass band amplitude.

## 3. Enhanced dynamic continualization for the wave propagation

To obtain an equivalent continuum model able to approximate finely the dynamic behavior of the Lagrangian one, an enhanced continualization approach proposed by Bacigalupo and Gambarotta, 2018, is here applied. A macroscopic generalized displacement field is considered, characterized by a non-dimensional displacement $\Psi(x,t)$ and rotation $\Phi(x,t)$. The derivative at point $x_i$ of such functions is assumed in terms of the central difference as follows

$$\left.\frac{\partial \Psi}{\partial x}\right|_{x_i} \doteq \frac{\psi_{i+1}-\psi_{i-1}}{2\ell} \quad , \quad \left.\frac{\partial \Phi}{\partial x}\right|_{x_i} \doteq \frac{\varphi_{i+1}-\varphi_{i-1}}{2\ell} \quad . \tag{9}$$

Furthermore, the shift operator $E_\ell$ is now taken into account to relate the displacement and rotation of two adjacent nodes $\psi_{i+1}=E_\ell \psi_i$ or $\varphi_{i+1}=E_\ell \varphi_i$ (see for references Jordan, 1965, Rota et al., 1973, Kelley and Peterson, 2001, among the others). Because of the notable property $E_\ell = \sum_{h=0}^{\infty} \frac{\ell^h}{h!} D^h = \exp(\ell D)$, with $D^h = \frac{\partial^h}{\partial x^h}$, the introduction of the particular pseudo-differential operator $\exp(\pm \ell D)$ in equation (9) can be demonstrated. Accordingly, the shift operation can be expressed as $\psi_{i\pm 1}(t) = \exp(\pm \ell D)\psi_i$ and the derivatives (9) may be written

$$\begin{aligned}\left.\frac{\partial \Psi}{\partial x}\right|_{x_i} &= D\Psi\big|_{x_i} = \left[\frac{\exp(\ell D)-\exp(-\ell D)}{2\ell}\right]\psi_i \quad , \\ \left.\frac{\partial \Phi}{\partial x}\right|_{x_i} &= D\Phi\big|_{x_i} = \left[\frac{\exp(\ell D)-\exp(-\ell D)}{2\ell}\right]\varphi_i \quad , \end{aligned} \tag{10}$$

where the terms between brackets are additional pseudo-differential operators. Therefore, the down-scaling laws for the block translation and rotation can be derived in terms of the continuous fields

$$\begin{aligned}\psi_i(t) &= \left[\frac{2\ell D}{\exp(\ell D)-\exp(-\ell D)}\right]\Psi(x,t)\bigg|_{x_i} \quad , \\ \varphi_i(t) &= \left[\frac{2\ell D}{\exp(\ell D)-\exp(-\ell D)}\right]\Phi(x,t)\bigg|_{x_i} \quad , \end{aligned} \tag{11}$$

where some notable mathematical properties of the pseudo-differential operators have been invoked to operate formally as with algebraic operators (Maslov, 1976, Shubin 1987).

By substituting equations (11) in the equations of motion (2), the system of two pseudo-differential equations is obtained

$$\begin{cases} P_1(D)\Psi - \ell D\Phi = I_\psi P_3(D)\ddot{\Psi} \\ \ell D\Psi + \dfrac{1}{12}\dfrac{k_n b^2}{k_t \ell^2} P_1(D)\Phi - \dfrac{1}{4} P_2(D)\Phi = \dfrac{I_\psi}{12}\left[1+\left(\dfrac{b}{\ell}\right)^2\right] P_3(D)\ddot{\Phi} \end{cases}, \quad (12)$$

involving the pseudo-differential operators

$$P_1(D) = \frac{2[\exp(\ell D) - 2 + \exp(-\ell D)]}{[\exp(\ell D) - \exp(-\ell D)]} \ell D ,$$

$$P_2(D) = \frac{2[\exp(\ell D) + 2 + \exp(-\ell D)]}{[\exp(\ell D) - \exp(-\ell D)]} \ell D ,$$

$$P_3(D) = \frac{2\ell D}{[\exp(\ell D) - \exp(-\ell D)]} .$$

In order to obtain the equations of motion of the equivalent homogeneous continuum, the McLaurin expansions of the pseudo-differential operators may be considered

$$\begin{aligned} P_1(D) &= \ell^2 D^2 - \frac{1}{12}\ell^4 D^4 + \frac{1}{120}\ell^6 D^6 + \mathcal{O}(\ell^7 D^7) , \\ P_2(D) &= 4 + \frac{1}{3}\ell^2 D^2 - \frac{1}{180}\ell^4 D^4 + \frac{1}{7560}\ell^6 D^6 + \mathcal{O}(\ell^7 D^7) , \\ P_3(D) &= 1 - \frac{1}{6}\ell^2 D^2 + \frac{7}{360}\ell^4 D^4 - \frac{31}{15120}\ell^6 D^6 + \mathcal{O}(\ell^7 D^7) , \end{aligned} \quad (13)$$

where $D$ is formally treatable as expansion variable. The expansions (13) can be substituted in (12) and the terms up to a selected order can be retained. If the terms up to the second order are considered the equations of motion of the second order equivalent continuum take the form

$$\begin{cases} \ell^2 \dfrac{\partial^2 \Psi}{\partial x^2} - \ell \dfrac{\partial \Phi}{\partial x} = I_\psi\left(\ddot{\Psi} - \dfrac{\ell^2}{6}\dfrac{\partial^2 \ddot{\Psi}}{\partial x^2}\right) \\ \ell \dfrac{\partial \Psi}{\partial x} - \Phi + \dfrac{1}{12}(r_k r_b^2 - 1)\ell^2 \dfrac{\partial^2 \Phi}{\partial x^2} = \dfrac{1}{12}I_\psi(1 + r_b^2)\left(\ddot{\Phi} - \dfrac{\ell^2}{6}\dfrac{\partial^2 \ddot{\Phi}}{\partial x^2}\right) \end{cases}. \quad (14)$$

These equations may be sought as the Euler-Lagrange equations corresponding to a Hamiltonian having the elastic potential energy density $\Pi_e = \dfrac{1}{2}\left[\left(\ell\dfrac{\partial \Psi}{\partial x} - \Phi\right)^2 + \dfrac{1}{12}(r_k r_b^2 - 1)\ell^2\left(\dfrac{\partial \Phi}{\partial x}\right)^2\right]$. The kinetic energy density may be shown to be unconditionally positive defined, while the elastic potential energy density of the model turns out to be positive defined for $r_k r_b^2 > 1$. Moreover, it is worth to note that the form of the elastic potential energy density is the same of the Timoshenko beam (Bhaskar, 2009, Wang and So, 2015), while the inertia terms in equation (14) and in the kinetic energy are here enriched with a non-local contribution depending on the second order space derivative of the acceleration field. This enhancement turns out to be necessary in consideration of the fact that both the acoustic and the optical branches from the Timoshenko model are always increasing with the wave number, a circumstance that may not take place in the Lagrangian model for several values of the geometric and constitutive parameters (see the acoustic spectra in Figure 2a,b,c). By assuming harmonic waves propagation in the form $\Psi(x,t) = \bar{\Psi} \exp[I(kx - \omega t)]$ and $\Phi(x,t) = \bar{\Phi} \exp[I(kx - \omega t)]$, the equations of motion specialize in the form

$$(\mathbf{H}_{Hom2} - \omega^2 \mathbf{I}_{Hom2}) \Upsilon = \mathbf{0} , \tag{15}$$

where $\mathbf{H}_{Hom2}$ is a Hermitian matrix, $\mathbf{I}_{Hom2}$ is a diagonal matrix assuming the form

$$\mathbf{H}_{Hom2} = \begin{bmatrix} k^2 \ell^2 & k\ell I \\ -k\ell I & 1 + \frac{1}{12}(r_k r_b^2 - 1)k^2 \ell^2 \end{bmatrix}$$

$$\mathbf{I}_{Hom2} = I_\psi \begin{bmatrix} 1 - \frac{1}{6}k^2 \ell^2 & 0 \\ 0 & \frac{1}{12}(1 + r_b^2)\left(1 - \frac{1}{6}k^2 \ell^2\right) \end{bmatrix} \tag{16}$$

and $\Upsilon^T = \{\bar{\Psi} \quad \bar{\Phi}\}$ is the macroscopic polarization vector. Again, two dispersion functions $\omega(k\ell)$ are obtained by solving the macroscopic spectral problem (15). Higher order approximations of the equations of motion can mathematically be obtained by retaining terms up to the fourth and sixth order, mechanically corresponding to higher order continualization in enhanced continua. Accordingly, the spectral problem (15) can be re-formulated and solved to determine the dispersion functions $\omega(k\ell)$ of the fourth order model and sixth order model.

A qualitative and quantitative comparison among the Floquet-Bloch spectra of the Lagrangian model and the three continuum formulations can be appreciated in Figure 6. As major remark, the comparison shows a good matching among the dispersion curves of the Lagrangian model (black lines) and the corresponding dispersion curves of the equivalent enhanced continua (blue, red, green lines). This mathematical result confirms that the enhanced continualization can consistently approximate the dispersion properties of the rigid stacked material described by the Lagrangian model. The approximation accuracy tends to decrease with the distance from the limit point of long wavelengths. The better approximation is systematically associated to the higher order continualizations.

As minor remark, although the positive definiteness of the elastic potential energy density can be lost for certain parameter combinations of the continuum models (for instance if $r_k r_b^2 \leq 1$ in the second order continuum model), nevertheless the optical branch is certainly positive. Differently, the Legendre-Hadamard ellipticity condition is respected as long as the acoustic branch is positive (together with the phase velocity). This positivity condition is actually satisfied in the majority of cases here considered, except in the spectra shown in Figure 6a and Figure 6b. Specifically, in the case shown in Figure 6a the positivity of the acoustic frequency is maintained up to a certain non-dimensional wave number ($k\ell = 0$ for the second order model, larger $k\ell$-values for both fourth and the sixth order model). Differently, in the case shown in Figure 6b only the acoustic frequency of the second order model is not positive, with the Legendre-Hadamard condition not satisfied.

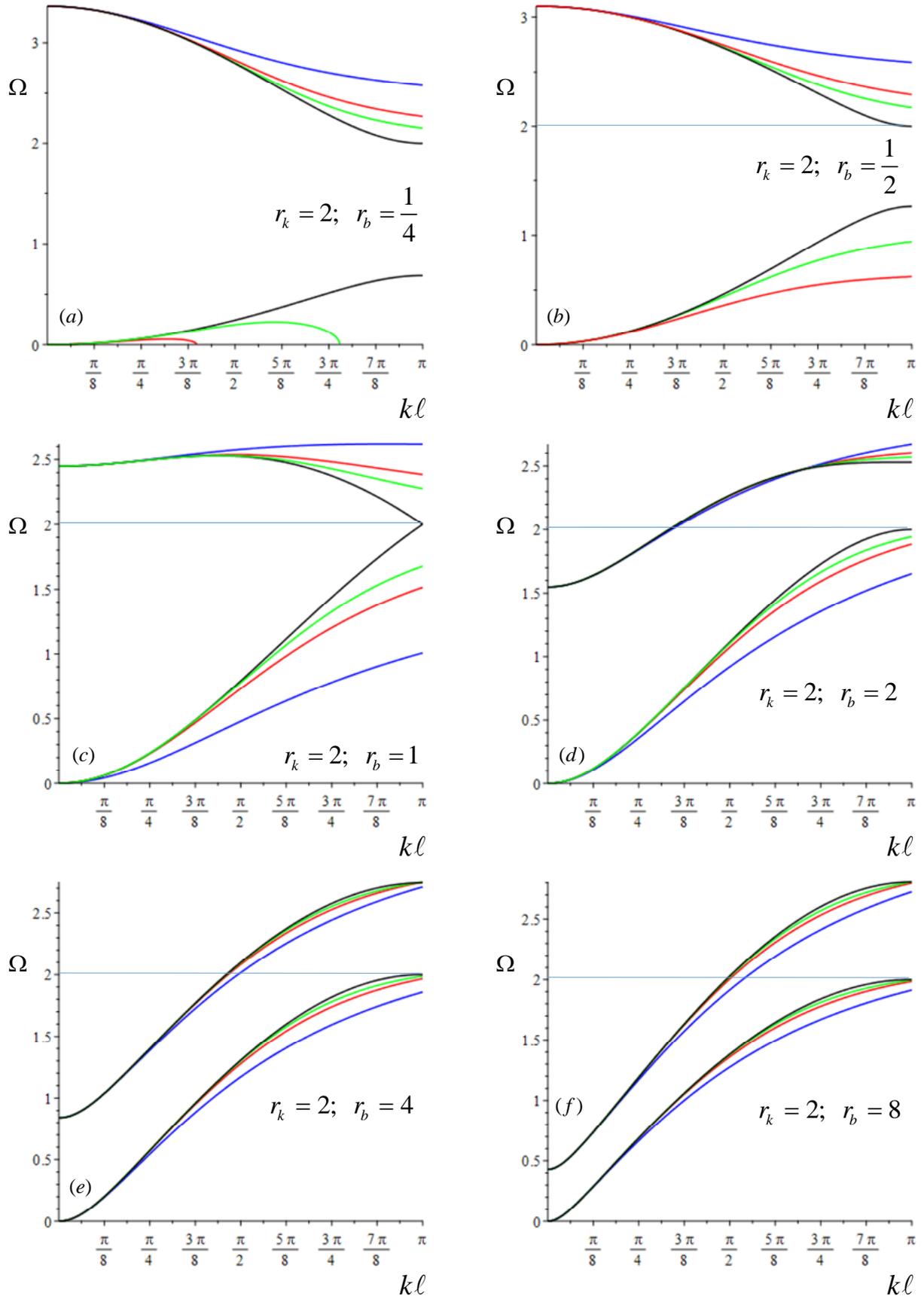

Fig. 6. Consistent approximation of the Floquet-Bloch spectra of the Lagrangian model (black line) via enhanced continualization: second order (blue line), fourth order (red line), sixth order (green line).

## 4. Wave propagation in a high performance high-pass acoustic filter made up of rigid stacked material with elastic interfaces

To improve the acoustic performances of the blocky system, including filtering of the ultra-low frequency waves, it may be useful to include an elastic half-space modelled as a Winkler support having stiffness $k_w$ and acting on the bottom side of the blocks, orthogonally to the elastic interfaces (see Figure 7).

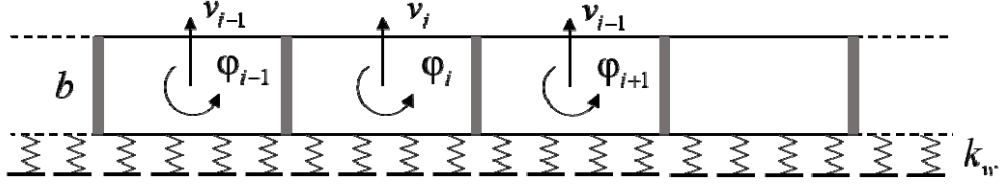

Fig. 7. Stack of rigid blocks on a continuous Winkler support.

The elastic restoring generalized forces acting on the $i$-th block have intensity $k_w v_i$ and $\frac{1}{12} k_w \ell^3 \varphi_i$ so that the equations of motion (2) are rewritten as

$$\begin{cases} \left[ \psi_{i+1} - \left(2 + \dfrac{r_{kw} r_k}{r_b}\right) \psi_i + \psi_{i-1} \right] - \dfrac{1}{2}(\varphi_{i+1} - \varphi_{i-1}) = I_\psi \ddot{\psi}_i \\ \dfrac{1}{2}(\psi_{i+1} - \psi_{i-1}) + \dfrac{1}{12} r_k r_b^2 \left[ \varphi_{i+1} - \left(2 + \dfrac{r_{kw}}{r_b^3}\right) \varphi_i + \varphi_{i-1} \right] + \\ -\dfrac{1}{4}(\varphi_{i+1} + 2\varphi_i + \varphi_{i-1}) = \dfrac{1}{12} I_\psi (1 + r_b^2) \ddot{\varphi}_i \end{cases} \qquad (17)$$

with the stiffness ratio $r_{kw} = \dfrac{k_w}{k_n}$. If the propagation of harmonic waves is considered, the equations of motion take the form

$$\left( \mathbf{H}^*_{Lag} - \omega^2 \mathbf{I}_{Lag} \right) \upsilon = \mathbf{0} , \qquad (18)$$

where $\mathbf{H}^*_{Lag}$ and $\mathbf{I}_{Lag}$ assume the form

$$\mathbf{H}^*_{Lag} = \begin{bmatrix} 2\left[\left(1+\dfrac{r_{kw}r_k}{2r_b}\right)-\cos(k\ell)\right] & I\sin(k\ell) \\ -I\sin(k\ell) & \dfrac{1}{2}\left\{\left[1+\dfrac{1}{3}r_k r_b^2\left(1+\dfrac{r_{kw}}{2r_b^3}\right)\right]+\left(1-\dfrac{1}{3}r_k r_b^2\right)\cos(k\ell)\right\} \end{bmatrix} \quad (19)$$

$$\mathbf{I}_{Lag} = I_\psi \begin{bmatrix} 1 & 0 \\ 0 & \dfrac{1}{12}(1+r_b^2) \end{bmatrix}$$

depending on the ratio $r_{kw}$. For the long wavelength approximation $(k\ell \to 0)$ the frequencies are $\Omega_1^0 = \sqrt{\dfrac{r_{kw} r_k}{r_b}}$ and $\Omega_2^0 = \sqrt{\dfrac{r_k r_{kw} + 12 r_b}{r_b(1+r_b^2)}}$, from which it appears that the stiffness of the lateral springs $r_{kw}$ modifies the qualitative structure of the Bloch spectra and can inhibit the acoustic wave propagation. In particular, the non-dimensional frequency $\Omega_1^0$ corresponds to a frequency $\omega_1^0 = \sqrt{\dfrac{k_w}{\rho b}}$, namely the frequency of lateral translation of the stack of rigid blocks in absence of relative generalized displacements between adjacent blocks. For short wavelength $(k\ell = \pi)$ the corresponding frequencies are $\Omega_1^\pi = \sqrt{\dfrac{r_{kw} r_k + 4 r_b}{r_b}}$ and $\Omega_2^\pi = \sqrt{\dfrac{r_k(r_{kw}+4r_b^3)}{r_b(1+r_b^2)}}$, respectively. It is worth to note that the occurrence $\Omega_1^0 = \Omega_2^0$ is attained for values of the constitutive ratio $r_{kw} = \dfrac{12}{r_b r_k}$, while $\Omega_1^\pi = \Omega_2^\pi$ is attained for values of the constitutive ratio $r_{kw} = 4\dfrac{(r_k - 1)r_b^2 - 1}{r_k r_b}$, a condition that may be obtained only for $r_b > \sqrt{\dfrac{1}{r_k - 1}}$.

The Floquet-Bloch spectra for different geometric ratios $r_b$ are shown in Figure 8 for a fixed realistic value of the constitutive ratio $r_k$ and different values of the stiffness ratio $r_{kw}$. The spectra of the freestanding periodic system ($r_{kw} = 0$) are taken as references (black lines). The effects of increasing stiffness ratios $r_{kw}$ are analyzed (blue, red, green lines). As major remark, increasing stiffness ratio determines a systematic increment of all the acoustic and optical frequencies. The consequent upshift of the acoustic branch causes the opening of a ultra-low frequency stop band. As minor remark, the stop bandwidth tends to increase for blocks with decreasing geometric ratio $r_b$ (see for instance Figure 8a versus Figure 8d). As complementary remarks, the amplitude of the stop band between the acoustic and optical branches is also modified.

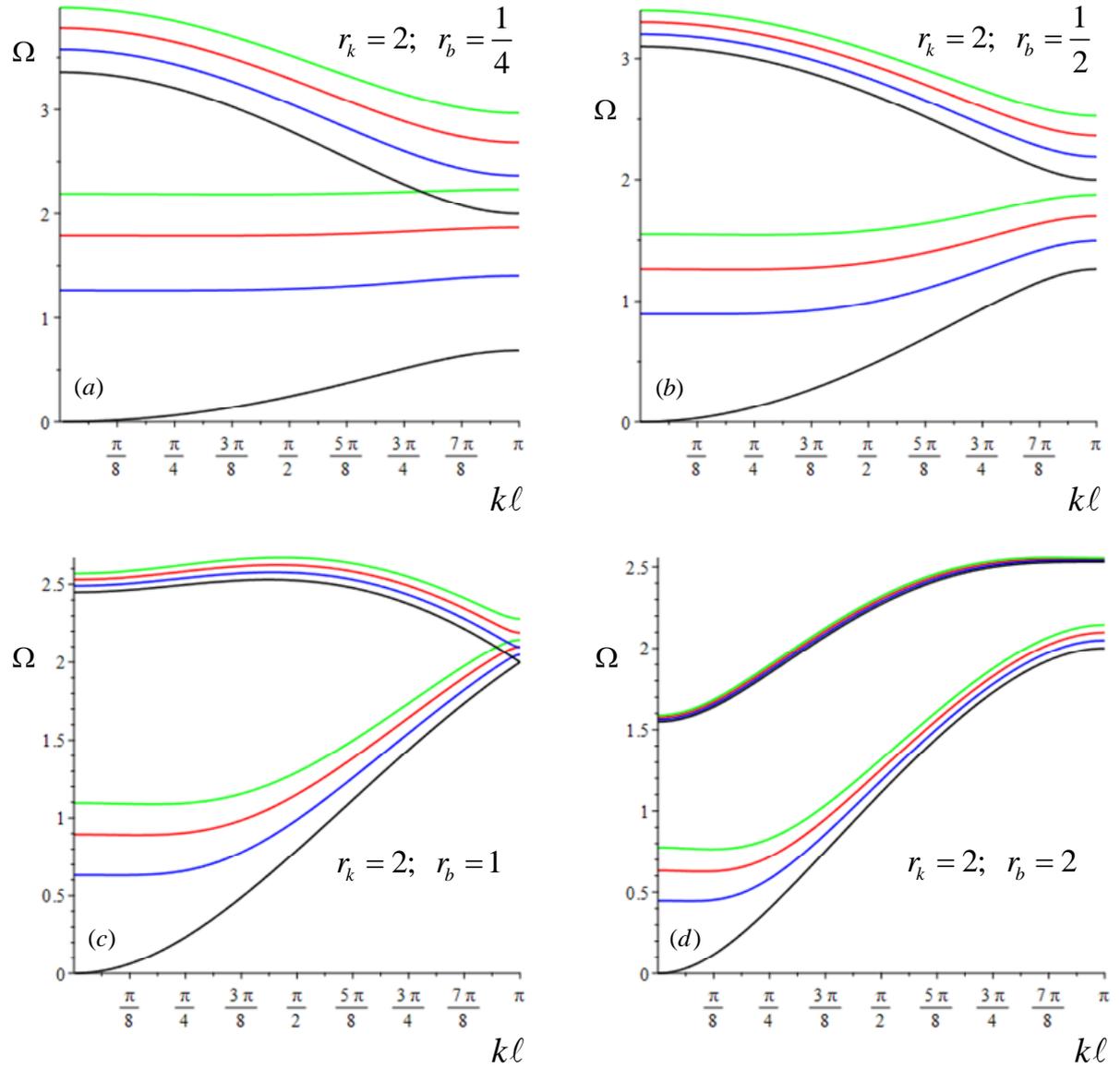

Fig. 8. Influence of the stiffness ratio $r_{kw}$ on the Floquet-Bloch spectra for $r_k = 2$ and different values of the geometric ratio $r_b$; stiffness ratio $r_{kw} = 0$ (black line), $r_{kw} = 1/5$ (blue line), $r_{kw} = 2/5$ (red line), $r_{kw} = 3/5$ (green line).

The equations of motion of the equivalent continuum are obtained according to the continualization procedure presented in Section 3. The simplest model is governed by the second order system of differential equation as follows

$$\begin{cases} -\dfrac{r_{kw}r_k}{r_b}\Psi + \left(1+\dfrac{r_{kw}r_k}{6r_b}\right)\ell^2\dfrac{\partial^2\Psi}{\partial x^2} - \ell\dfrac{\partial\Phi}{\partial x} = I_\psi\left[\ddot\Psi - \dfrac{\ell^2}{6}\dfrac{\partial^2\ddot\Psi}{\partial x^2}\right] \\ \ell\dfrac{\partial\Psi}{\partial x} - \left(1+\dfrac{1}{12}\dfrac{r_k r_{kw}}{r_b}\right)\Phi + \dfrac{1}{12}\left[r_k r_b^2\left(1+\dfrac{r_{kw}}{6r_b^3}\right)-1\right]\ell^2\dfrac{\partial^2\Phi}{\partial x^2} = \\ \qquad = \dfrac{1}{12}I_\psi\left[1+\left(\dfrac{b}{\ell}\right)^2\right]\left[\ddot\Phi - \dfrac{\ell^2}{6}\dfrac{\partial^2\ddot\Phi}{\partial x^2}\right] \end{cases} \qquad (20)$$

It is worth to note that the condition for the positive definiteness of the elastic potential energy turns out to be $r_k r_b^2\left(1+\dfrac{r_{kw}}{6r_b^3}\right) > 1$.

The circular frequencies and the wave vectors are obtained in terms of the non-dimensional wave number $k\ell$ by solving the eigenvalue problem

$$\left(\mathbf{H}^*_{Hom2} - \omega^2 \mathbf{I}^*_{Hom2}\right)\Upsilon = \mathbf{0}, \qquad (21)$$

where $\mathbf{H}^*_{Hom2}$ and $\mathbf{I}^*_{Hom2}$ assume the form

$$\mathbf{H}^*_{Hom2} = \begin{bmatrix} \left[\dfrac{r_{kw}r_k}{r_b} + \left(1+\dfrac{r_{kw}r_k}{6r_b}\right)k^2\ell^2\right] & Ik\ell \\ -Ik\ell & \left\{\left(1+\dfrac{1}{12}\dfrac{r_k r_{kw}}{r_b}\right) + \dfrac{1}{12}\left[r_k r_b^2\left(1+\dfrac{r_{kw}}{6r_b^3}\right)-1\right]k^2\ell^2\right\} \end{bmatrix} \qquad (22)$$

$$\mathbf{I}^*_{Hom2} = I_\psi \begin{bmatrix} \left(1+\dfrac{1}{6}k^2\ell^2\right) & 0 \\ 0 & \dfrac{1}{12}\left[1+\left(\dfrac{b}{\ell}\right)^2\right]\left(1+\dfrac{1}{6}k^2\ell^2\right) \end{bmatrix}$$

whose solutions coincide, in the long wavelength approximation $k\ell \to 0$ with those of equation (18). Higher order enhanced and standard continualizations are given in Appendix A and Appendix B, respectively.

The capability of the dispersion functions given by equation (22) to simulate those of the Lagrangian model, is synthetically shown in the diagrams of Figure 9. As major remark, the comparison shows a good approximation among the dispersion curves of the Lagrangian model (black lines) and the corresponding dispersion curves of the equivalent enhanced continua (blue, red, green lines). Again, the approximation accuracy tends to decrease with the distance from the limit point of long wavelengths. Higher order continualizations are required to achieve better approximation.

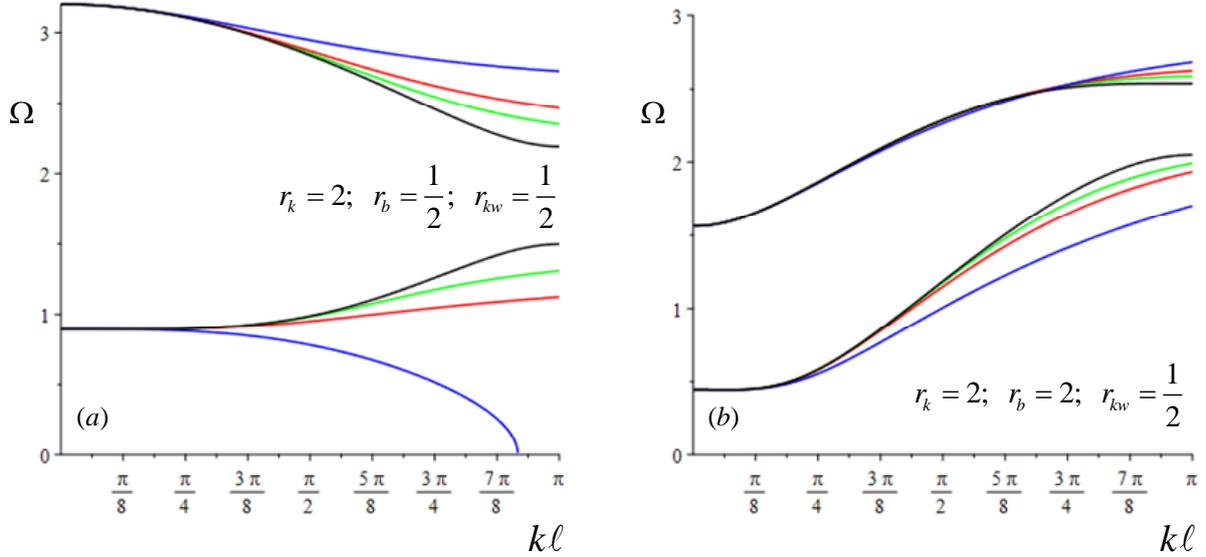

Fig. 9. Consistent approximation of the Floquet-Bloch spectra of the Lagrangian model (black line) via enhanced continualization: second order (blue line), fourth order (red line), sixth order (green line).

## 6. Conclusions

Rigid stacked materials made of periodic massive blocks connected to each other by linear elastic joints can be considered simple prototypical mechanical systems to describe the acoustic dispersion properties of monodimensional waveguide filters. Specifically, the filtering performance of the stacked material characterized by a periodic cell with two degrees of freedom (transversal displacement and rotation) has been parametrically analysed. To this purpose, the linear eigenproblem governing the wave propagation of a Lagrangian model has been solved analytically to achieve closed form dispersion relations. The dispersion spectrum is systematically featured by an acoustic low-frequency branch and an optical high-frequency branch, corresponding respectively to waveforms dominated by the shear and moment components at the limit of long wavelengths. Considering an optimal design perspective, the frequency amplitudes of the spectral pass and stop bands have been conveniently assessed in the bi-dimensional space of the independent mechanical parameters. In order to achieve ultra-low frequency stop bands without the introduction of super-massive and/or extra-flexible local resonators, the elastic coupling with an elastic half-space modelled as a Winkler support is considered. As major achievement, the dispersion spectrum turns out to be characterized by two optical branches, due to the systematic upshift of the entire band structure. Furthermore, the lower-frequency branch develops a critical point (corresponding to vanishing group velocity) with not-null frequency at the limit of long wavelengths. The consequent stop bands opened in the ultra-low frequency range has been described in the enlarged (three-dimensional) space of the independent mechanical parameters.

Moving from the spectral characterization of the Lagrangian models, a convenient fine approximation of the dispersion relations has been pursued by formulating homogenised continuum models, suited for macroscopic descriptions of the dynamic response even in finite complex-shaped domains. In order to overcome some shortcomings inherent to the classic homogenization of periodic blocky materials, an enhanced continualization approach has been employed. Specifically, the enhanced continualization allows identifying homogenised models featured by higher order micropolar formulations, which provide good descriptions of the acoustic behaviour, without incurring – in the general case – in the drawback of a non-positive definite elastic potential energy density. The mathematical treatment essentially consists in transforming the difference equations of motion governing the Lagrangian model into a pseudo-differential problem, by means of a proper down-scaling law relating the block degrees of freedom to continuous field variables. Therefore, partial differential equations of motion are obtained by adopting a proper McLaurin approximation of the governing pseudo-differential operator. Higher order continuum formulations turn out to be governed by the equations of motion achievable by including more terms of the truncated McLaurin series. As major remark from the physical viewpoint, the enhanced continualization introduces non-local inertia terms with spatial high gradients. The dispersion relations of the homogenised continua are eventually found to closely approximate the entire spectrum (acoustic and optical branches) of the Lagrangian models both in the absence and in the presence of the supporting elastic half-space.

**Acknowledgements**

The authors acknowledge financial support of the (MURST) Italian Department for University and Scientific and Technological Research in the framework of the research MIUR Prin15 project 2015LYYXA8, Multi-scale mechanical models for the design and optimization of micro-structured smart materials and metamaterials, coordinated by prof. A. Corigliano. The authors also thankfully acknowledge financial support by National Group of Mathematical Physics (GNFMINdAM).

**Compliance with ethical standards**

The authors declare that they have no conflict of interest

**References**

Acar, G., Yilmaz, C., Experimental and numerical evidence for the existence of wide and deep phononic gaps induced by inertial amplification in two-dimensional solid structures. *Journal of Sound and Vibration*, **332**, 6389-6404, 2013.

Bacigalupo A., Gambarotta L., Dispersive wave propagation in two-dimensional rigid periodic blocky materials with elastic interfaces, *Journal of the Mechanics and Physics of Solids,* **102**, 165-186, 2017.

Bacigalupo, A., Gambarotta, L., Generalized micropolar continualization of 1D beam lattices, submitted, 2018. *https://arxiv.org/abs/1808.07125*.

Baravelli, E., Ruzzene, M., Internally resonating lattices for bandgap generation and low-frequency vibration control. *Journal of Sound and Vibration*, **332**, 6562-6579, 2013.

Beli, D., Arruda, J. R. F., Ruzzene, M., Wave propagation in elastic metamaterial beams and plates with interconnected resonators. *International Journal of Solids and Structures*, **139**, 105-120, 2018.

Bhaskar A., Elastic waves in Timoshenko beams: the 'lost and found' of an eigenmode, *Proc. Royal Society A*, **465**, 239-255, 2009.

Brillouin, L., *Wave Propagation in Periodic Structures*, McGraw-Hill, New York, 1946.

Brun, M., Giaccu, G.F., Movchan, A.B., Movchan, N.V., Asymptotics of eigenfrequencies in the dynamic response of elongated multi-structures, *Proc. R. Soc. A*, **468**, 378-394, 2012.

Carta, G., Jones, I.S., Movchan, N.V., Movchan, A.B., Nieves, M.J., Gyro-elastic beams for the vibration reduction of long flexural systems. *Proc. R. Soc. A*, **473**, 20170136, 2017.

Chen Y., Wang L., Bio-inspired heterogeneous composites for broadband vibration mitigation, *Scientific Reports Nature*, **5**, 17865, 2015.

Craster, R.V., Guenneau,S. (Eds.), *Acoustic Metamaterials*, Springer-Verlag, London, 2012.

Cummer S.A., Christensen J., Alù A., Controlling sound with acoustic metamaterials, Nature Reviews, Materials, **1**, 1-13, 2016.

Deng, B., Wang, P., He, Q., Tournat, V., Bertoldi, K., Metamaterials with amplitude gaps for elastic solitons. *Nature communications*, **9**, 3410, 2018.

Deymier, P.A. (Ed.), *Acoustic metamaterials and phononic crystals* (Vol. 173). Springer Science & Business Media, 2013.

Frandsen, N.M., Bilal, O.R., Jensen, J.S., Hussein, M.I., Inertial amplification of continuous structures: Large band gaps from small masses, *Journal of Applied Physics*, **119**(12), 124902, 2016.

Gei, M., Movchan, A.B., Bigoni, D., Band-gap shift and defect-induced annihilation in prestressed elastic structures, *Journal of Applied Physics*, **105**, 063507, 2009.

Habermann M.R., Norris A.N., Acoustic metamaterials, *Acoustic Today*, **12**, 31-39, 2016.

Huang, G. L., Sun, C. T., Band gaps in a multiresonator acoustic metamaterial. *Journal of Vibration and Acoustics*, **132**, 031003, 2010.

Hussein, M. I., Leamy, M. J., Ruzzene, M. Dynamics of phononic materials and structures: Historical origins, recent progress, and future outlook. *Applied Mechanics Reviews*, **66**(4), 040802, 2014.

Jordan C., *Calculus of finite differences*, Vol. 33. American Mathematical Soc., 1965.

Kelley W.G., Peterson A.C., *Difference equations. An introduction with applications*, Academic Press, 2001.

Ma G., Sheng P., Acoustic metamaterials: From local resonances to broad horizons, S*cience Advances*, **2**, 1501595, 2016.

Maslov, V. P., *Operational methods,* Mir, 1976.

Matlack, K.H., Serra-Garcia, M., Palermo, A., Huber, S.D., Daraio, C., Designing perturbative metamaterials from discrete models. *Nature materials*, **17**, 323, 2018.

Miniaci, M., Krushynska, A., Movchan, A. B., Bosia, F., Pugno, N.M. Spider web-inspired acoustic metamaterials. *Applied Physics Letters*, **109**(7), 071905, 2016.

Miniaci, M., Krushynska, A., Gliozzi, A.S., Kherraz, N., Bosia, F., Pugno, N.M., Design and fabrication of bioinspired hierarchical dissipative elastic metamaterials. *Physical Review Applied*, **10**(2), 024012, 2018.

Piccolroaz, A., Movchan, A.B., Dispersion and localisation in structured Rayleigh beams, *International Journal of Solids and Structures*, **51**, 4452-4461, 2014.

Rota G.C., Kahaner D., Odlyzko A., On the foundations of combinatorial theory. VIII. Finite operator calculus. *Journal of Mathematical Analysis and Applications,* 42, 684-760, 1973.

Shubin, M.A., *Pseudodifferential operators and spectral theory,* Berlin: Springer-Verlag, 1987.

Taniker, S., Yilmaz, C., Design, analysis and experimental investigation of three-dimensional structures with inertial amplification induced vibration stop bands, *International Journal of Solids and Structures*, **72**, 88-97, 2015.

Vasseur, J.O., Deymier, P.A., Chenni, B., Djafari-Rouhani, B., Dobrzynski, L., Prevost, D., Experimental and theoretical evidence for the existence of absolute acoustic band gaps in two-dimensional solid phononic crystals. *Physical Review Letters*, **86**, 3012, 2001.

Wang, X.Q., So, R.M.C., Timoshenko beam theory: A perspective based on the wave-mechanics approach. *Wave Motion*, **57**, 64-87, 2015.

Yilmaz, C., Hulbert, G.M., Kikuchi, N., Phononic band gaps induced by inertial amplification in periodic media, *Physical Review B*, **76**, 054309, 2007.

Yin, J., Huang, J., Zhang, S., Zhang, H. W., Chen, B. S., Ultrawide low frequency band gap of phononic crystal in nacreous composite material, *Physics Letters A*, **378**, 2436-2442, 2014.

Yin, J., Peng, H. J., Zhang, S., Zhang, H. W., Chen, B. S., Design of nacreous composite material for vibration isolation based on band gap manipulation. *Computational Materials Science*, **102**, 126-134, 2015.

**Appendix A – The sixth order PDE governing the equivalent homogenized model**

By applying the enhanced continualization procedure presented in Section 3 to the system of equations (17) governing the motion of the block, a system of two PDEs is obtained, whose order depends on the order of truncation applied. Here, the equations derived by retaining terms up to the sixth order are given, having the following form

$$\begin{cases} -\dfrac{r_{kw}r_k}{r_b}\Psi + \left(1+\dfrac{r_{kw}r_k}{6r_b}\right)\ell^2\dfrac{\partial^2\Psi}{\partial x^2} - \left(\dfrac{1}{12}+\dfrac{7}{360}\dfrac{r_{kw}r_k}{r_b}\right)\ell^4\dfrac{\partial^4\Psi}{\partial x^4} + \\ +\left(\dfrac{1}{120}+\dfrac{31}{15120}\dfrac{r_{kw}r_k}{r_b}\right)\ell^6\dfrac{\partial^6\Psi}{\partial x^6} - \ell\dfrac{\partial\Phi}{\partial x} = \\ \quad = I_\psi\left(\ddot{\Psi} - \dfrac{\ell^2}{6}\dfrac{\partial^2\ddot{\Psi}}{\partial x^2} + \dfrac{7\ell^4}{360}\dfrac{\partial^4\ddot{\Psi}}{\partial x^4} - \dfrac{31\ell^6}{15120}\dfrac{\partial^6\ddot{\Psi}}{\partial x^6}\right) \\ \ell\dfrac{\partial\Psi}{\partial x} - \left(1+\dfrac{1}{12}\dfrac{r_k r_{kw}}{r_b}\right)\Phi + \dfrac{1}{12}\left[r_k r_b^2\left(1+\dfrac{r_{kw}}{6r_b^3}\right)-1\right]\ell^2\dfrac{\partial^2\Phi}{\partial x^2} + \\ -\dfrac{1}{144}\left[r_k r_b^2\left(1+\dfrac{7}{30}\dfrac{r_{kw}}{r_b^3}\right)-\dfrac{1}{60}\right]\ell^4\dfrac{\partial^4\Phi}{\partial x^4} + \\ +\dfrac{1}{1440}\left[r_k r_b^2\left(1+\dfrac{31}{126}\dfrac{r_{kw}}{r_b^3}\right)-\dfrac{1}{21}\right]\ell^6\dfrac{\partial^6\Phi}{\partial x^6} = \\ \quad = \dfrac{1}{12}I_\psi\left[1+\left(\dfrac{b}{\ell}\right)^2\right]\left(\ddot{\Phi} - \dfrac{\ell^2}{6}\dfrac{\partial^2\ddot{\Phi}}{\partial x^2} + \dfrac{7\ell^4}{360}\dfrac{\partial^4\ddot{\Phi}}{\partial x^4} - \dfrac{31\ell^6}{15120}\dfrac{\partial^6\ddot{\Phi}}{\partial x^6}\right). \end{cases} \quad (A.1)$$

Lower order formulations are derived retaining terms of fourth and second order, the latter being given in equation (20). It is worth to note that also for the higher order model the condition for the positive definiteness of the elastic potential energy turns out to be $r_k r_b^2\left(1+\dfrac{r_{kw}}{6r_b^3}\right) > 1$.

**Appendix B – Standard continualization**

Let us consider a standard continualization of (17) through the introduction of the shift operator previously introduced. The pseudo-differential problem of equilibrium for the *i*-th block takes the form

$$\begin{cases} \left[\exp(\ell D) - \left(2 + \dfrac{r_{kw} r_k}{r_b}\right) + \exp(-\ell D)\right]\psi_i + \\ -\dfrac{1}{2}\left[\exp(\ell D) - \exp(-\ell D)\right]\varphi_i = I_\psi \ddot{\psi}_i \\ \dfrac{1}{2}\left[\exp(\ell D) - \exp(-\ell D)\right]\psi_i + \\ +\dfrac{1}{12} r_k r_b^2 \left[\exp(\ell D) - \left(2 + \dfrac{r_{kw}}{r_b^3}\right) + \exp(-\ell D)\right]\varphi_i + \\ -\dfrac{1}{4}\left[\exp(\ell D) + 2 + \exp(-\ell D)\right]\varphi_i = \dfrac{1}{12} I_\psi \left(1 + r_b^2\right)\ddot{\varphi}_i \end{cases} \quad . \tag{B.1}$$

By expanding in a Taylor series the pseudo-differential operators a PDE system is obtained. In case the expansion is truncated to the fourth order one obtains

$$\begin{cases} -\dfrac{r_{kw} r_k}{r_b}\Psi + \ell^2 \dfrac{\partial^2 \Psi}{\partial x^2} + \dfrac{1}{12}\ell^4 \dfrac{\partial^4 \Psi}{\partial x^4} - \ell \dfrac{\partial \Phi}{\partial x} - \dfrac{2}{3}\ell^3 \dfrac{\partial^3 \Phi}{\partial x^3} = I_\psi \dfrac{\partial^2 \Psi}{\partial t^2} \\ \ell \dfrac{\partial \Psi}{\partial x} + \dfrac{2}{3}\ell^3 \dfrac{\partial^3 \Psi}{\partial x^3} - \left(1 + \dfrac{1}{12}\dfrac{r_k r_{kw}}{r_b}\right)\Phi + \dfrac{1}{12}\left(r_k r_b^2 - 3\right)\ell^2 \dfrac{\partial^2 \Phi}{\partial x^2} \\ +\dfrac{1}{144}\left(r_k r_b^2 - 3\right)\ell^4 \dfrac{\partial^4 \Phi}{\partial x^4} = \dfrac{1}{12} I_\psi \left(1 + r_b^2\right)\dfrac{\partial^2 \Phi}{\partial t^2} \end{cases} \quad , \tag{B.2}$$

and the corresponding elastic potential energy density turns out to be

$$\Pi_e = \dfrac{1}{2}\left[\begin{array}{l} \dfrac{1}{6}\dfrac{r_k r_{kw}}{r_b}\Phi^2 + \left(\ell \dfrac{\partial \Psi}{\partial x} - \Phi\right)^2 + \dfrac{1}{6}\left(r_k r_b^2 - 5\right)\left(\ell \dfrac{\partial \Phi}{\partial x}\right)^2 + \\ +\dfrac{1}{6}\ell^2 \left[\dfrac{\partial}{\partial x}\left(\ell \dfrac{\partial \Psi}{\partial x} - \Phi\right)\right]^2 + \dfrac{1}{6}\left(\ell^2 \dfrac{\partial^2 \Psi}{\partial x^2}\right)^2 + \\ +\dfrac{1}{72}\left(r_k r_b^2 - 3\right)\left(\ell^2 \dfrac{\partial^2 \Phi}{\partial x^2}\right)^2 \end{array}\right] \quad . \tag{B.3}$$

The elastic potential energy density turns out to be positive defined for values of the ratios such that $r_k r_b^2 > 5$, which is more restrictive than the corresponding one obtained through the enhanced homogenization $r_k r_b^2 \left(1 + \dfrac{r_{kw}}{6 r_b^3}\right) > 1$ related to the differential problem (20).